\documentclass[doublecol]{epl2}
\usepackage{amsmath}
\usepackage{amsfonts}
\usepackage{graphicx}
\usepackage{amssymb}
\usepackage{float}

\title{Dynamics of a Rydberg hydrogen atom near a topologically insulating surface}
\author{A. Mart\'{i}n-Ruiz\inst{1,2} and E. Chan-L\'{o}pez\inst{3}} 

\institute{
\inst{1} Instituto de Ciencia de Materiales de Madrid, CSIC, Cantoblanco, 28049 Madrid, Spain \\ 
\inst{2} Instituto de Ciencias Nucleares, Universidad Nacional Aut\'{o}noma de M\'{e}xico, 04510 M\'{e}xico D. F., M\'{e}xico \\
\inst{3} Universidad Ju\'{a}rez Aut\'{o}noma de Tabasco, DACB, 86690 Cunduac\'{a}n, Tabasco, M\'{e}xico}


\abstract{We investigate the classical dynamics of a Rydberg hydrogen atom near the surface of a planar topological insulator. The system is described by a Hamiltonian consisting of the free-hydrogen part and the hydrogen-surface potential. The latter includes the interactions between the electron and both image electric charges and image magnetic monopoles. Owing to the axial symmetry, the $z$ component of angular momentum $l _{z}$ is conserved. Here we consider the $l _{z} = 0$ case. The structure of the phase space is explored extensively by means of numerical techniques and Poincar\'{e} surfaces of section for the recently discovered topological insulator TlBiSe$_{2}$. The phase space of the system is separated into regions of vibrational and rotational motion. We show that vibrational-rotational-vibrational type transitions can be tuned with the topological magnetoelectric polarizability.}

\begin{document}

\maketitle

\section{Introduction} \label{Introduction}
The interaction between atoms and surfaces is of fundamental importance in the fields of physics, chemistry and biology. For example, atom-surface interactions play an important role in atomic force microscopy \cite{Binnig} and they also affect the dynamical properties of an atom or molecule nearby. Hence the considerable amount of attention it has gained in the last decades.

The study of electron dynamics and spectroscopy of highly excited Rydberg atoms in external fields have been a very active field of research in recent years \cite{Inarrea1, Inarrea2, Hua1, Salas}. Rydberg atoms are particularly important in these studies since they are sensitive to perturbations due to their large size and the weak binding of the excited electron; besides, they can be prepared and manipulated in the laboratory. The study of the dynamics of Rydberg atoms near a metallic surface have attracted great attention since this system can simulate many dynamics effects of atoms in strong fields, such as the Zeeman-Stark effect, diamagnetic effect and instantaneous van der Waals interaction, for example. Since the interaction of a Rydberg atom with a metal surface takes place far from the surface (compared with the atom size), the atom-surface interaction can be modeled by the electrostatic method of images, i.e. the images of the electric charges of the atom act as another atom \cite{Ganesan, Simonovic, Dunning, Simonovic2}. Therefore, the image atom exerts additional forces on the atomic electron, thus affecting its dynamical properties. This van der Waals force between the atom and the nearby metal surface plays a vital role in the adsorption process. The same idea has been extended to Rydberg atoms near dielectric surfaces, finding a chaotic or regular classical motion, depending on the atom-surface distance \cite{Hua2, Hua3, Hua4}. In this letter we aim to study the dynamics of a Rydberg atom near a topologically insulating surface.

Topological insulators (TIs) are an emerging class of materials which have attracted much attention in condensed matter physics. They are characterized by a bulk insulating behavior with metallic surface states protected by time-reversal (TR) symmetry \cite{Qi-Review, Liang}. In addition to their interesting electronic properties, TIs also display unusual electromagnetic properties. Specifically, the topological magnetoelectric effect (TME), which consist in the mixing of the electric and magnetic induction fields. Many interesting TME have been predicted but none of them have been detected in the laboratory. The most striking consequence of the TME is the image magnetic monopole effect \cite{Qi-Science}, which consists in the following. Consider bringing an electric charge near the surface of a TI, in addition to the image electric charge an image magnetic monopole will also appear inside the material. Physically, the monopole magnetic field is induced by a circulating vortex Hall current on the surface of the TI, which is sourced by the (tangential component of the) electric field next to the interface, rather than by a point magnetic charge.

The problem we shall consider in this letter is that of bringing a Rydberg hydrogen atom near the surface of a TI, as shown in fig. \ref{System}. Due to the image magnetic monopole effect, the electric charges of the atom will produce image electric charges and image magnetic monopoles located inside the material, which in turn will interact with the atomic electron. More precisely, here we study the dynamics of the atomic electron when interacting with the image electric and magnetic fields. Using numerical techniques and Poincar\'{e} surfaces of section, we explore extensively the structure of the phase space for the TI TlBiSe$_{2}$. The phase space of the system is found to be separable into regions of vibrational and rotational motion. We show that vibrational-rotational-vibrational type transitions can be tuned with the topological magnetoelectric polarizability (TMEP) $\theta$.

\section{Rydberg hydrogen atom near a TI surface} In this section, we briefly review the electromagnetic response of TIs and the image magnetic monopole effect, to then establish the Hamiltonian of the system.

\subsection{Electromagnetic response of TIs}
\begin{figure}[tbp]
\begin{center}
\includegraphics[width=2.5in]{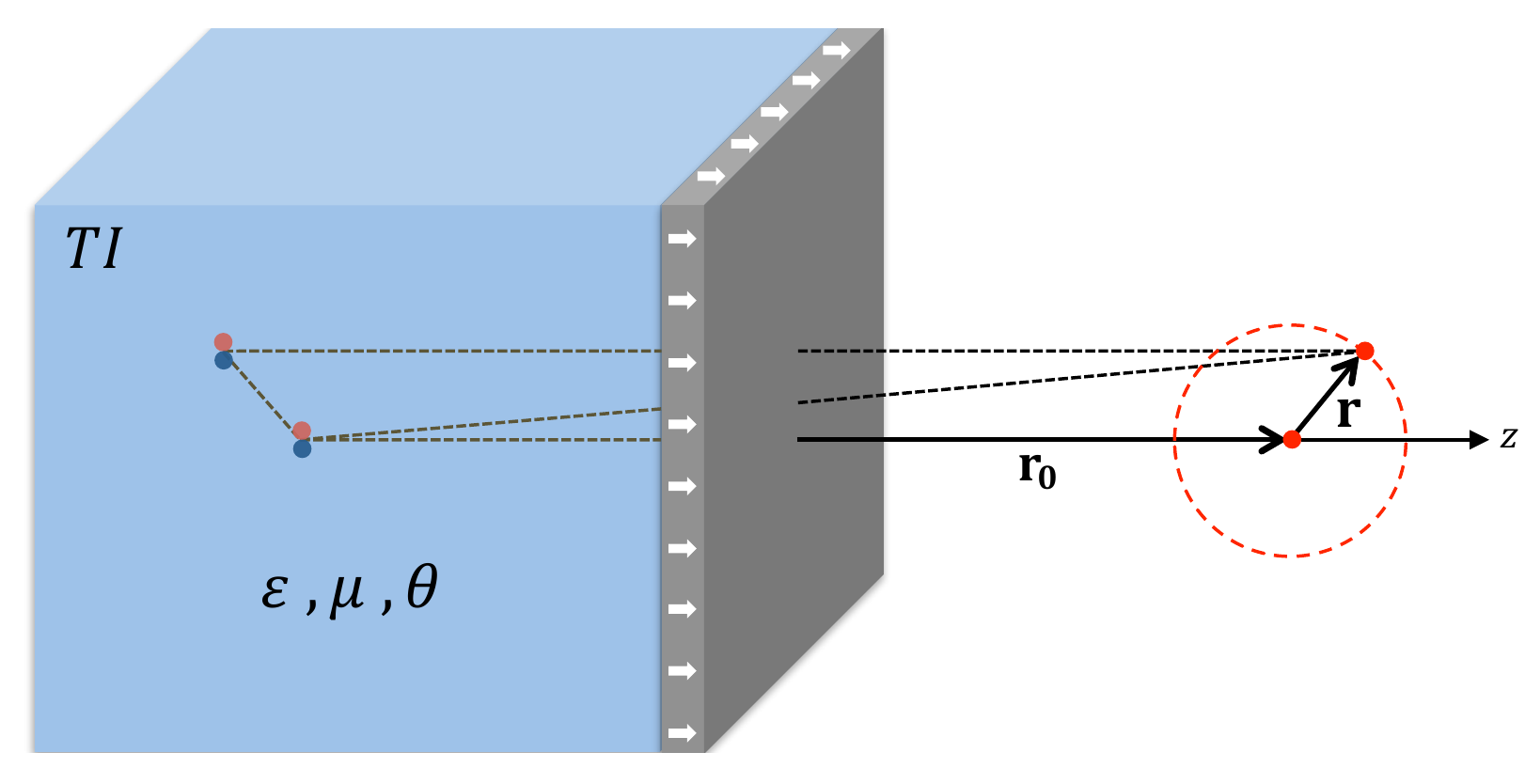}
\end{center}
\caption{{\protect\small Schematic of a Rydberg hydrogen atom near a three-dimensional TI half-space and its image electric charges and image magnetic monopoles. The TI is covered with a thin magnetic layer (not to scale) which gaps the surface states.}}
\label{System}
\end{figure}

It has been suggested recently, based on subtle field-theoretical considerations, that the low-energy effective field theory which describes the electromagnetic response of TIs (independently of microscopic details) is defined by the Lagrangian

\begin{align}
\mathcal{L} = \frac{1}{8 \pi} \left( \varepsilon \textbf{E} ^{2} - \frac{1}{\mu} \textbf{B} ^{2} \right) + \frac{\alpha}{4 \pi ^{2}} \theta \textbf{E} \cdot \textbf{B} , \label{Lagrangian}
\end{align}
where $\textbf{E}$ and $\textbf{B}$ are the electromagnetic fields, $\alpha \simeq 1 / 137$ is the fine structure constant, $\varepsilon$ and $\mu$ are the permittivity and permeability, respectively, and $\theta$ is the TMEP. Because of TR symmetry, the last term is a good description of the bulk of a trivial insulator when $\theta = 0$ and the bulk of a TI when $\theta = \pi$.cWhen the boundary is included, this theory is a fair description of both the bulk and the surface only when a TR breaking perturbation is induced on the surface to gap the surface states. In this work we consider that the TR perturbation is a magnetic coating of small thickness which gaps the surface fermions, as shown in fig. \ref{System}. In the described situation, for a TI hosting $2n+1$ surface fermions (with $n \in \mathbb{Z}$), $\theta$ can be shown to be quantized as $\theta = \pm ( 2n+1 ) \pi$. Positive and negative values of $\theta$ are related to different signs of the magnetization in the direction perpendicular to the surface.

The axion coupling term in Eq. (\ref{Lagrangian}) does not modify Maxwell equations with redefined constituent relations $\textbf{D} = \varepsilon \textbf{E} + \alpha (\theta / \pi) \textbf{B}$ and $\textbf{H} = \textbf{B} / \mu - \alpha (\theta / \pi) \textbf{E}$. Therefore, nontrivial effects due to the topological term (known as TME) appear only at the interface of a TI and trivial insulator, where the TMEP suddenly changes. Specific TMEs have been predicted. For example, when polarized light propagates through a TI surface, of which the surface states has been gapped by TR symmetry breaking, topological Faraday and Kerr rotations take place \cite{Maciejko}. It was also recently proposed that the sign of the Casimir force between TIs can be tuned by means of the TMEP \cite{Grushin1, Grushin2}. As we will see later, the dynamics of a Rydberg hydrogen atom near a TI can also be tuned by means of the TMEP in a similar fashion.

\subsection{Image magnetic monopole effect}

Through the constitutive relations and the usual electromagnetic equations, an applied electric field can induce magnetization, while a magnetic field can induce polarization. More precisely, a tangential electric field on the surface of a TI would generate a transverse surface current, giving rise to a half-integer quantum Hall effect. Therefore, when a pointlike electric charge is brought near to the surface of a TI, the tangential component of the electric field induces a vortex Hall current, which generates a magnetic field that can be simulated by an image magnetic monopole inside the material. This is known as the image magnetic monopole effect.

This image magnetic monopole effect can be studied straightforwardly in the same way as the image charge problem in an ordinary insulator \cite{Qi-Science}. However, the same result can also be obtained using far superior techniques, such as the $SL (2 , \mathbb{Z})$ electric-magnetic duality group of TIs \cite{Karch} and Green's function techniques \cite{MCU1, MCU2, MCU3, MCU4}. In short, let us consider the geometry of fig. \ref{System}. The left-half space ($z<0$) is occupied by a TI with dielectric constant $\varepsilon$, magnetic permeability $\mu$ and TMEP $\theta$, whereas the right-half space ($z>0$) is the vacuum. A point electric charge $q$ is located in vacuum at $\textbf{r} _{0} = b \hat{\textbf{e}} _{z}$ from the TI. For $z>0$, the electric field can be interpreted as due to two point electric charges, one of strength $q$ at $\textbf{r} _{0}$, and the other, the image charge, of strength $q ^{\prime} = - \kappa q$, with

\begin{align}
\kappa = \frac{(\varepsilon - 1) (1 + 1 / \mu) + \tilde{\alpha} ^{2}}{(\varepsilon + 1) (1 + 1 / \mu) + \tilde{\alpha} ^{2}} \quad , \quad \tilde{\alpha} = \alpha (\theta / \pi ) , \label{kappa}
\end{align}
at the mirror point $- \textbf{r} _{0}$. The magnetic field can be interpreted as that of a magnetic monopole of strength

\begin{align}
g = \frac{2 q \tilde{\alpha}}{(\varepsilon + 1) (1 + 1 / \mu) + \tilde{\alpha} ^{2}} , \label{ImageMonopole}
\end{align}
at $- \textbf{r} _{0}$. Going back to our problem, when a Rydberg hydrogen atom (composed by two point electric charges) is brought near to the surface of a TI, as shown in fig. \ref{System}, in addition to the image electric charges (showed in red), image magnetic monopoles also appear inside the material (showed in blue). The interaction of these images with the atomic electron will affect the dynamical properties of the latter. This is the chief motivation of this letter.

\subsection{Hamiltonian}
Here we derive the Hamiltonian of the classical system depicted in fig. \ref{System}. Let us consider the motion of an electron in a Coulomb field induced by an infinitely massive nucleus of charge $e >0$ at the origin of the coordinate system. The TI-vacuum interface is located at $z = -b$, and the TI is assumed to be covered with a thin magnetic coating such that the surface states are gapped and thus the TME takes place. Therefore, the atomic electron will be affected by image electric charges and image magnetic monopoles as well, and thus the appropriate Hamiltonian is that of a charged particle in electromagnetic fields, i.e. $\mathcal{H} =  \left( \textbf{p} + e \textbf{A} \right) ^{2} / 2 m _{e} - e \phi (\textbf{r})$. The interaction between the involved electric charges and the atomic electron is given by

\begin{align}
- e \phi (\textbf{r}) = - \frac{e ^{2}}{r} + \frac{\kappa e ^{2}}{\sqrt{\rho ^{2} + (z+2b) ^{2}}} - \frac{\kappa e ^{2}}{4 (z + b)} , \label{CoulombInt}
\end{align}
where $\rho ^{2} = x ^{2} + y ^{2}$ and $r ^{2} = \rho ^{2} + z ^{2}$.  The first term describes the usual Coulomb interaction between the nucleus and the atomic electron, and the last two terms account for the interaction between the latter and the two image electric charges.

Due to the TME, the atomic charges will also produce image magnetic monopoles, whose magnetic fields will in turn interact with the atomic electron. In this letter we are concerned with the classical picture of this system, and therefore the electron spin is not considered. However, the image monopole magnetic fields will affect the orbital motion. In the Coulomb gauge, the vector potential is

\begin{align}
\textbf{A} (\textbf{r}) = \frac{g}{\rho ^{2}} \left( 1 - \frac{z + 2 b}{\sqrt{\rho ^{2} + (z + 2 b) ^{2}}} \right) \left( y \hat{\textbf{e}} _{x} - x \hat{\textbf{e}} _{y} \right) . \label{VectorInt}
\end{align}
This is nothing but the Schwinger vector potential, which is singular along the Dirac string $\vartheta = \pi$ \cite{Schwinger}. The last two terms in the Coulomb interaction (\ref{CoulombInt}) and the vector potential (\ref{VectorInt}) break the spherical symmetry of the free atom. However, the axial symmetry (around the $z$ axis) is still present. Therefore, in cylindrical coordinates $(\rho,  z, \phi , p _{\rho} , p _{z} , p _{\phi})$ and atomic units, the Hamiltonian $\mathcal{H}$ of the system reads

\begin{align}
\mathcal{H} = \frac{p ^{2} _{\rho} + p ^{2} _{z}}{2} + \frac{p _{\phi} ^{2}}{2 \rho ^{2}} + \textbf{p} \cdot \textbf{A} + \frac{\textbf{A} ^{2}}{2} - \frac{1}{r} \notag \\ + \frac{\kappa}{\sqrt{\rho ^{2} + (z+2b) ^{2}}} - \frac{\kappa}{4 (z + b)} . \label{Hamiltonian}
\end{align}
Owing to the axial symmetry, the $z$ component $l _{z} = p _{\phi}$ of the angular momentum is conserved and the system (\ref{Hamiltonian}) is reduced to a dynamical system with two degrees of freedom. Here we consider the $l _{z} = 0$ case, in such a way that, besides the energy $E = \mathcal{H}$, the dynamical behavior of the atomic electron will depend on the external parameters $\kappa$, $g$ and $b$.

\section{Classical dynamics of a Rydberg hydrogen atom near a TI surface}

The hydrogen atom is one of the rare real physical systems which is integrable. In the presence of external fields, its integrability depends on the type of perturbation. Since the perturbations in eq. (\ref{Hamiltonian}) couples the $\rho$ and $z$-coordinates, our problem is in general non-integrable. Therefore, here we explore the phase space structure by means of numerical techniques and Poincar\'{e} surfaces of section for the TI TlBiSe$_{2}$

\subsection{Scaling and regularization}
To proceed with the numerical analysis of the dynamical behaviour of the system, we scale the coordinates and momentum in the form: $\tilde{\textbf{r}} = \textbf{r} / b$, $\tilde{\textbf{p}} = \textbf{p} \sqrt{b}$. After dropping tildes, Hamiltonian (\ref{Hamiltonian}) becomes

\begin{align}
\tilde{\mathcal{H}} \equiv \xi = \frac{p ^{2} _{\rho} + p ^{2} _{z}}{2} - \frac{1}{\sqrt{\rho ^{2} + z ^{2}}} + \frac{\kappa}{\sqrt{\rho ^{2} + (z +2) ^{2}}} \notag \\ - \frac{\kappa}{4 (z + 1)} + \frac{\tilde{g} ^{2}}{2 \rho ^{2}} \left( 1 - \frac{z + 2}{\sqrt{\rho ^{2} + (z +2) ^{2}}} \right) ^{2} , \label{ScaledHamiltonian}
\end{align}
and the dynamics does not depend on the energy $E$ and the nucleus-surface distance $b$ separately, but only on the scaled energy $\xi = E b$. Also, we observe that the $\kappa$-parameter is unaffected by the scaling; however, the magnetic monopole strength becomes a running coupling constant $\tilde{g} = g / \sqrt{b}$ which depends on the distance $b$.

It is useful to study the shape of the effective potential $\mathcal{U} (\rho ,z) = U (\rho ,z) + V (\rho ,z)$ in eq. (\ref{ScaledHamiltonian}),

\begin{align}
U (\rho ,z) &= - \frac{1}{\sqrt{\rho ^{2} + z ^{2}}} + \frac{\kappa}{\sqrt{\rho ^{2} + (z +2) ^{2}}} - \frac{\kappa}{4 (z + 1)} , \notag \\ V (\rho ,z) &= \frac{\tilde{g} ^{2}}{2 \rho ^{2}} \left( 1 - \frac{z + 2}{\sqrt{\rho ^{2} + (z +2) ^{2}}} \right) ^{2} ,
\end{align}
through the determination of its critical points $P _{c} = (\rho _{c} , z _{c})$, which are given by the roots of the equations $\partial _{z} \mathcal{U} =0$ and $\partial _{\rho} \mathcal{U} = 0$. On the one hand, we observe that due to the Coulombic term $U (\rho ,z)$, the effective potential exhibits an infinite potential well at the origin. On the other hand, we find that the function $V (\rho , z)$ arising from the magnetic monopole contribution satisfies the following properties: $\lim _{\rho \rightarrow 0} V = 0$, $\lim _{\rho \rightarrow 0} \partial _{z} V = 0$ and $\lim _{\rho \rightarrow 0} \partial _{\rho} V = 0$. Therefore, the critical points of $\mathcal{U} (\rho ,z)$, when they exist, take place on the $z$ axis ($\rho _{c} = 0$). One can directly verify that the condition $\partial _{\rho} \mathcal{U} = 0$ is satisfied at $\rho _{c}$. Therefore we determine $z _{c}$ from

\begin{align}
\partial _{z} \mathcal{U} (\rho _{c} , z _{c}) = \frac{z _{c}}{\vert z _{c} \vert ^{3}} - \frac{\kappa}{(z _{c} + 2) ^{2}} + \frac{\kappa}{4 (z _{c} +1) ^{2}} = 0 . \label{ConditionZ}
\end{align}
Note that $V (\rho , z)$ does not contribute to the determination of the critical point since it is singular along the Dirac string ($z$ axis). Furthermore, due to the divergent electron-(image)electron interaction at $z=-1$, a critical point always exist in the interval $(-1,0)$. To proceed with the numerical determination of $z _{c}$, we fix constants by choosing the recently discovered TI TlBiSe$_{2}$ \cite{TlBiSe2, TlBiSe22, TlBiSe23}, for which $\theta = \pi$, $\mu = 1$ and $\varepsilon = 4$, then $\kappa \approx 0.6$. Numerical solution of eq. (\ref{ConditionZ}) in the region of interest yields $z _{c} = - 0.739304$. By substituting the critical point $P _{c} = (0, z _{c})$ in the corresponding Hessian matrix we readily obtain that $P _{c}$ is a saddle point whose energy is $\xi _{c} = -1.45208$.

Physically, because $l _{z} = 0$ there is no centrifugal barrier and the electron can reach the origin, where the Hamiltonian (\ref{ScaledHamiltonian}) presents a singularity. The saddle point $P _{c}$ is the ionization channel through which the atomic electron can be captured by the surface. To avoid the Coulomb singularity, we perform the so-called Levi-Civita regularization \cite{LeviCivita}. This procedure consists in a change to semiparabolic coordinates $(u,v)$,

\begin{align}
\rho = uv \qquad , \qquad z = (u ^{2} - v ^{2}) / 2 . 
\end{align}
The conjugate momenta, $p _{u} = du / d \tau$ and $p _{v} = dv / d \tau$ are defined with respect to the scaled time $\tau \equiv t / (u ^{2} + v ^{2})$. Finally, after an overall multiplication by $u ^{2} + v ^{2}$, the Hamiltonian (\ref{ScaledHamiltonian}) reads

\begin{align}
\tilde{\mathcal{H}} ^{\prime} &= 2 = \frac{p _{u} ^{2} + p _{v} ^{2}}{2} - \xi (u ^{2} + v ^{2}) \label{ParabolicHamiltonian} \\ &  + \frac{2 \kappa (u ^{2} + v ^{2})}{\sqrt{4 u ^{2} v ^{2} + (u ^{2} - v ^{2} + 4) ^{2}}} - \frac{\kappa (u ^{2} + v ^{2})}{2 ( u ^{2} - v ^{2} + 2)} \notag \\  & + \frac{\tilde{g} ^{2}}{2} \left( \frac{1}{u ^{2}} + \frac{1}{v ^{2}} \right) \left( 1 - \frac{u ^{2} - v ^{2} + 4}{\sqrt{4 u ^{2} v ^{2} +  (u ^{2} - v ^{2} + 4) ^{2}}} \right) ^{2} . \notag 
\end{align}
Note that the regularized Hamiltonian $\tilde{\mathcal{H}} ^{\prime}$ takes a constant value $2$ and the scaled energy $\xi$ appears as a parameter. The Hamilton equations of motion arising from (\ref{ParabolicHamiltonian}) are

\begin{align}
\dot{u} = p _{u} \qquad &, \qquad \dot{v} = p _{v} , \\ \dot{p} _{u} = - \frac{\partial \tilde{\mathcal{H}} ^{\prime}}{\partial u } \qquad &, \qquad \dot{p} _{v} = - \frac{\partial \tilde{\mathcal{H}} ^{\prime}}{\partial v} . 
\end{align}
Next we explore the structure of the phase space by means of numerical techniques and Poincar\'{e} surfaces of section.

\begin{figure}[tbp]
\begin{center}
\includegraphics[width=1.7in]{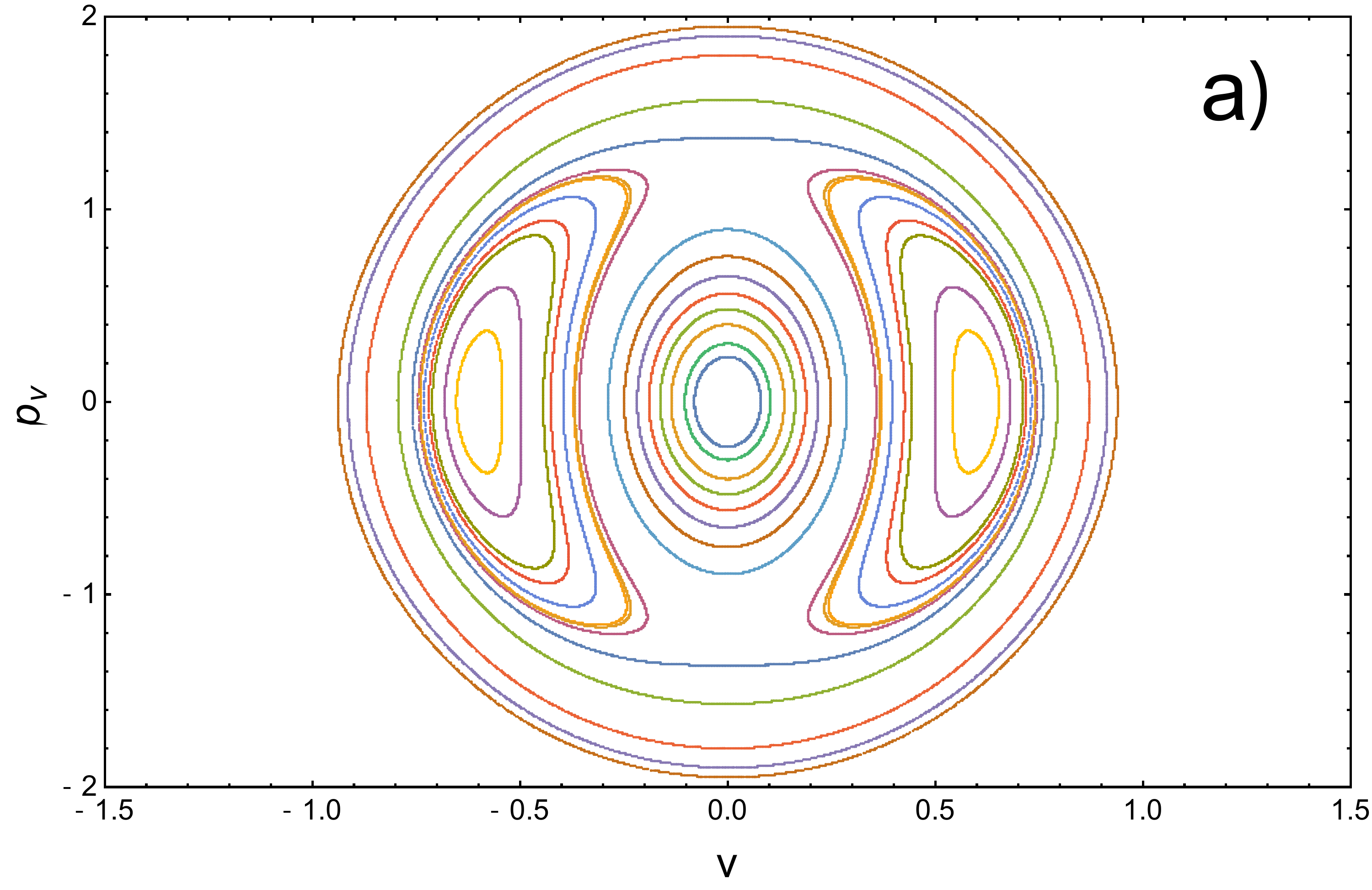} \label{PS1}
\includegraphics[width=1.7in]{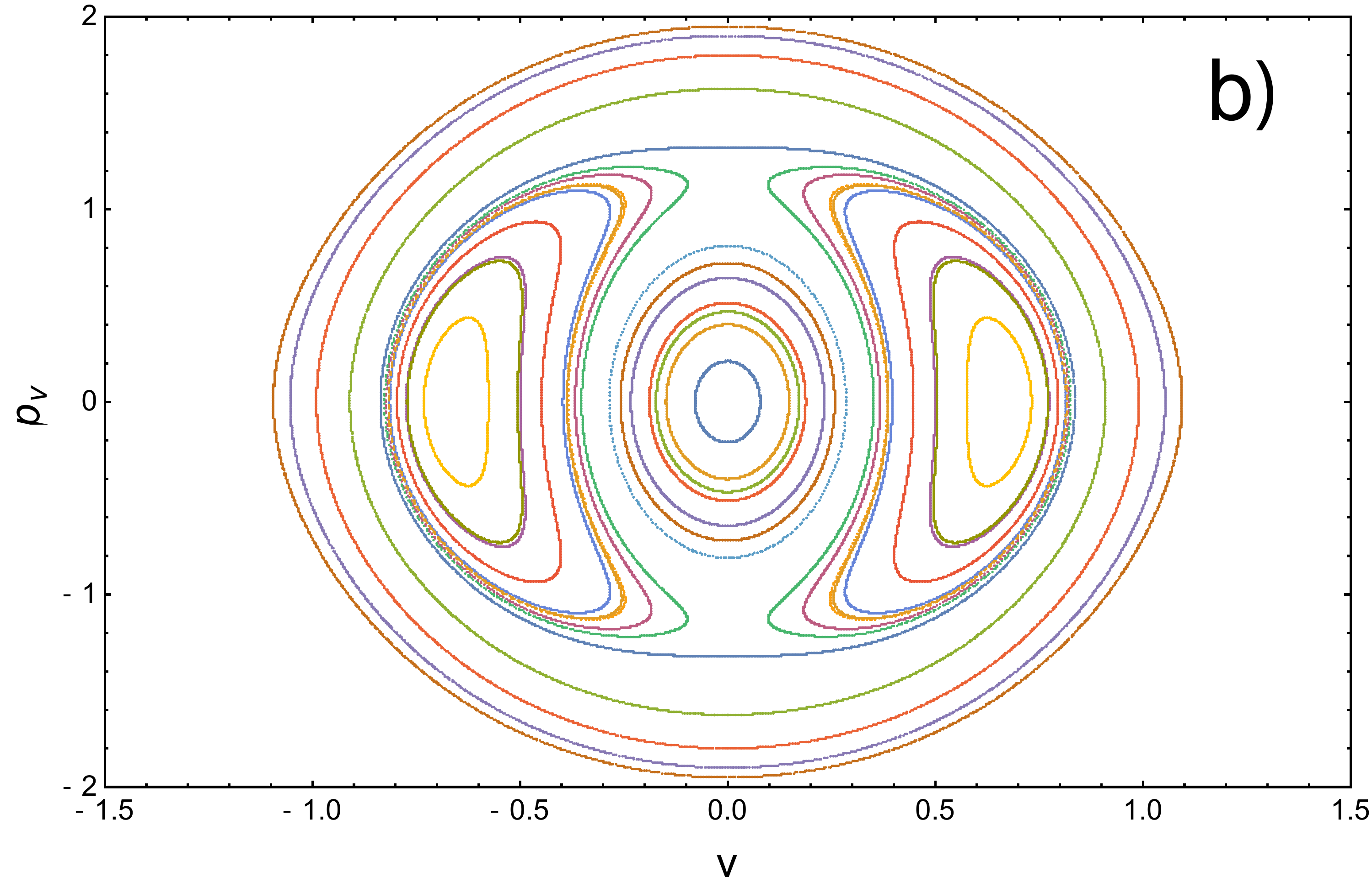} \label{PS2} \\
\includegraphics[width=1.7in]{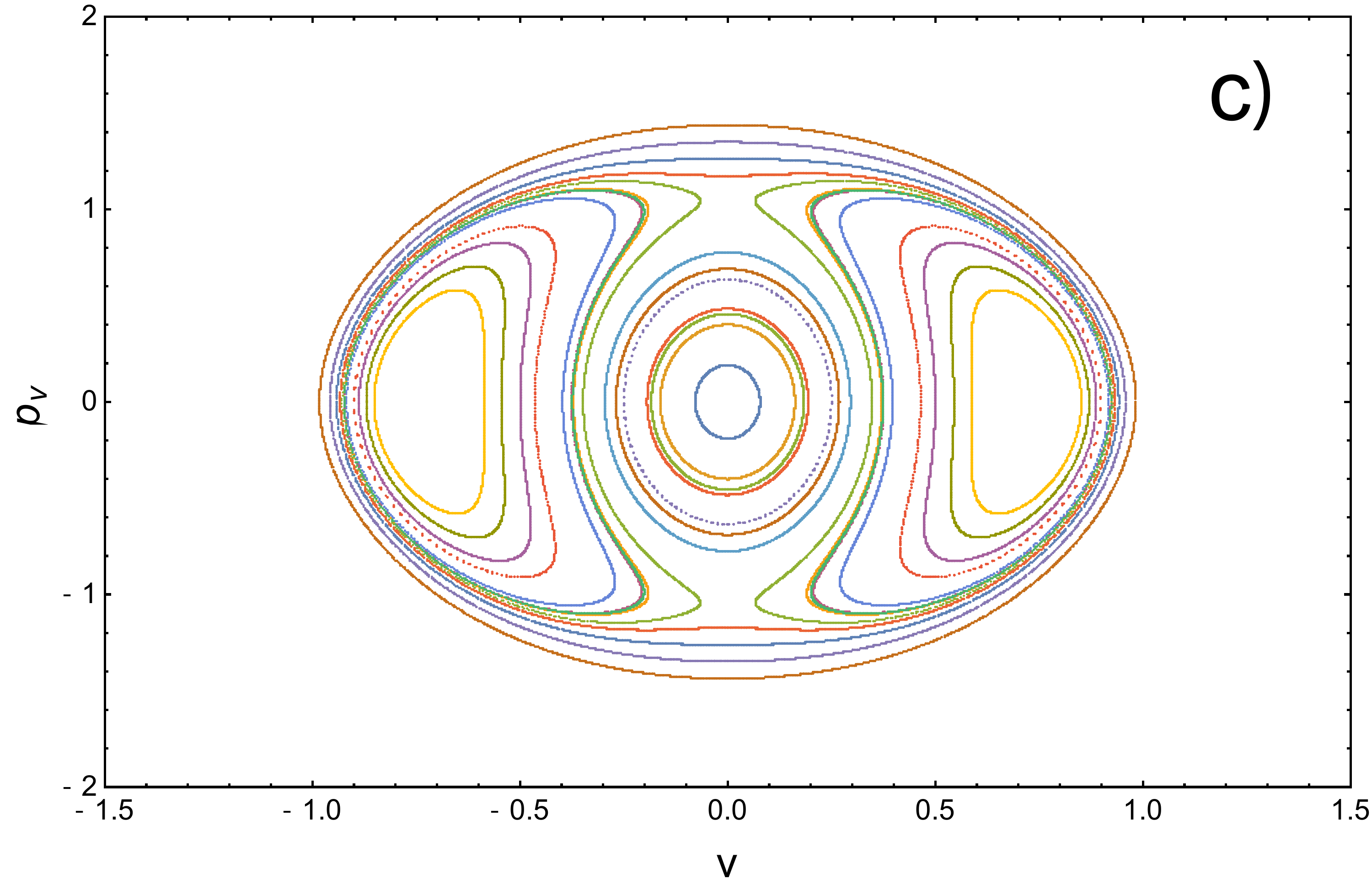} \label{PS3}
\includegraphics[width=1.7in]{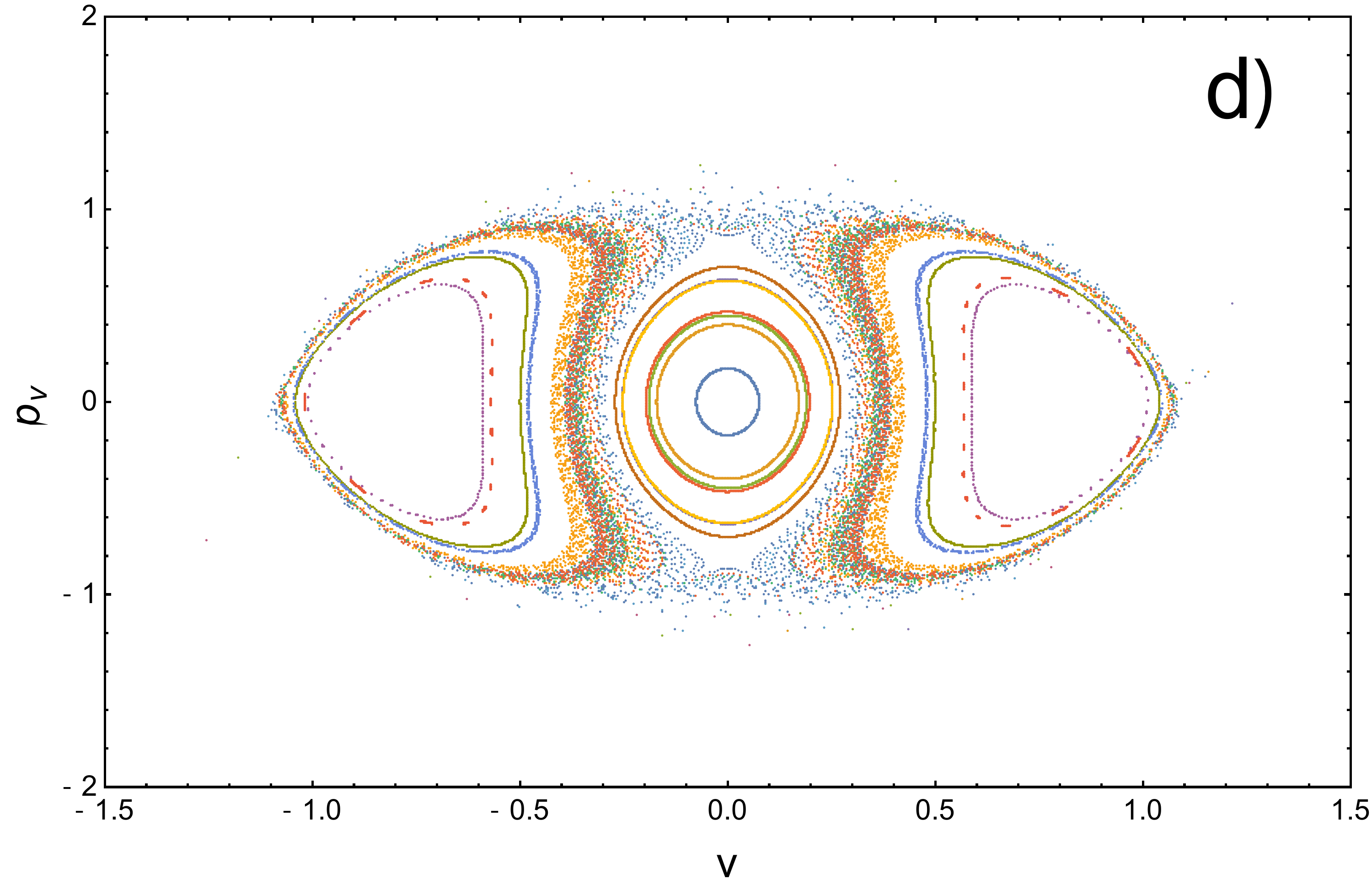} \label{PS4}
\end{center}
\caption{{\protect\small Poincar\'{e} surfaces of section for different values of the scaled energy}.}
\label{Poincare}
\end{figure}

\subsection{Phase space structure}

To get an overall vision of the phase space structure, we fix the nucleus-surface distance to be $b = 10, 000$ a.u. $\approx 530$ nm, which is appropriate for experimental realization with Rydberg hydrogen atoms. In fig. \ref{Poincare}, we plot the Poincar\'{e} surfaces of section for different values of the scaled energy $\xi$. The Poincar\'{e} surface of section is plotted in the $v-p_{v}$ plane; therefore, it is defined by all trajectories which intersect $u = 0$ with $p _{u} >0$. 

\begin{figure}[tbp]
\begin{center}
\includegraphics[width=1.7in]{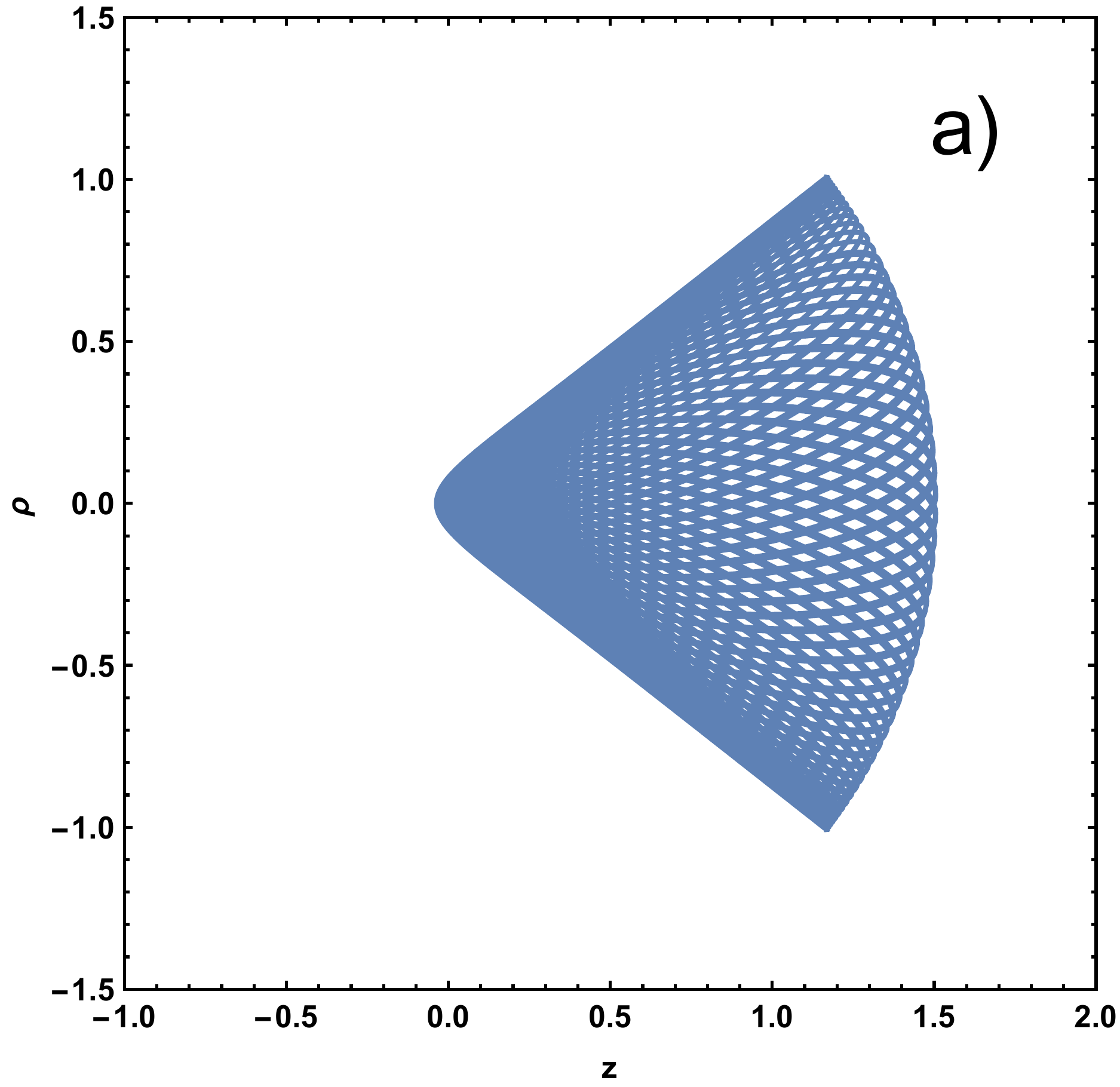} \label{Vib}
\includegraphics[width=1.7in]{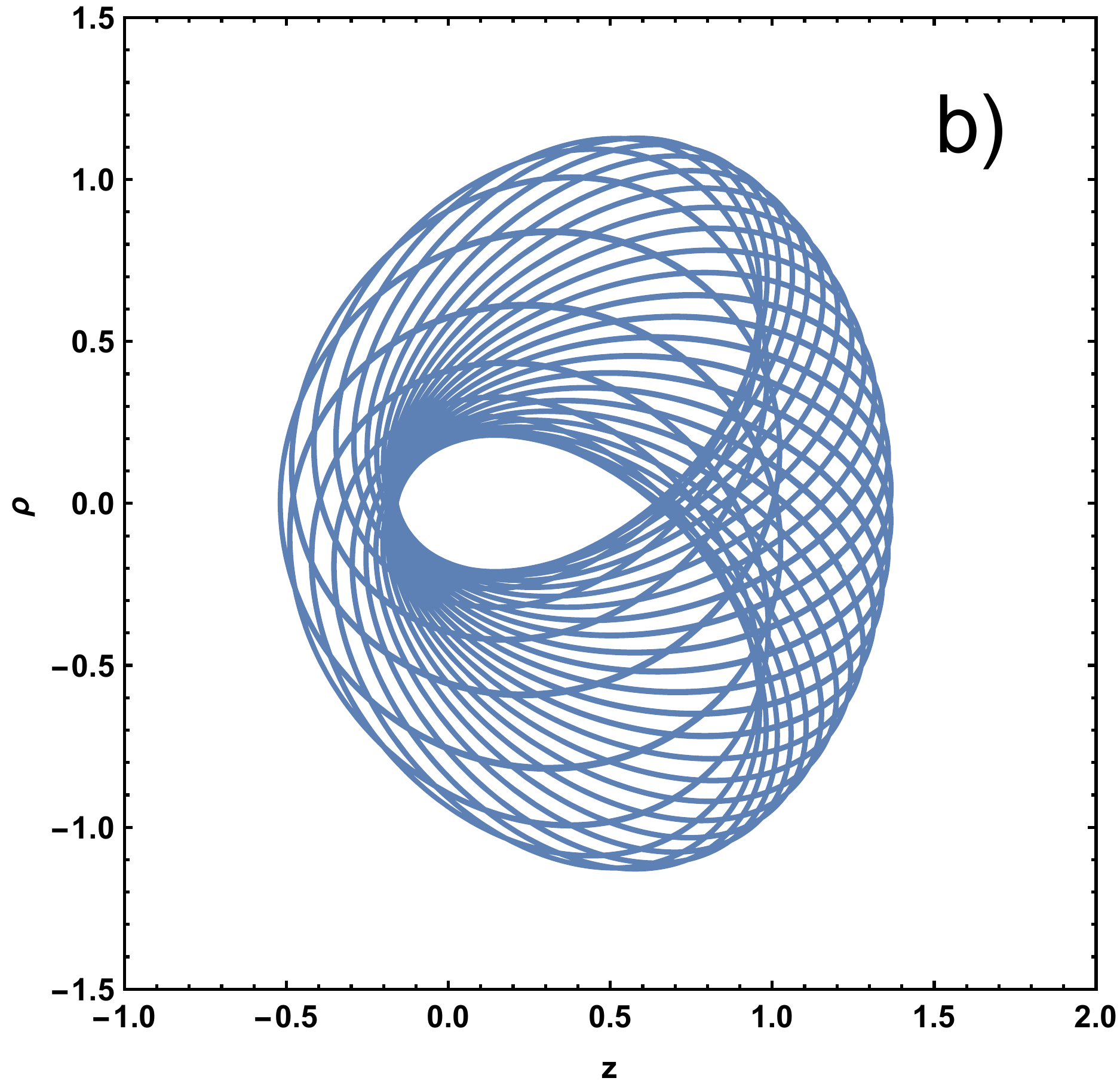} \label{Rot}
\end{center}
\caption{{\protect\small Examples for: a) vibrational and b) rotational types of orbits.}}
\label{Orbits}
\end{figure}

There is an essential difference between the cases $\xi < \xi _{c}$ and $\xi > \xi _{c}$, where the critical value is $\xi _{c} = -1.45208$, which corresponds to a highly excited atom with the principal quantum number $n = 59$. We first discuss the case $\xi < \xi _{c}$, where the electron is confined into the infinite potential well and its dynamics is still close to the integrable limit $\xi \rightarrow - \infty$. Figures \ref{Poincare}a and \ref{Poincare}b show the Poincar\'{e} surfaces of section for $\xi = -2$ ($n = 50 $) and $\xi = -1.5$ ($n = 58$), respectively. We observe that the phase space is divided in several regions. The stable fixed point in the central region $(v=0,p _{v} = 0)$ corresponds to rectilinear orbits along the $u$ axis, which corresponds to physical orbits along the positive $z$ axis. The ellipses around this fixed point are vibrational-type quasiperiodic orbits with the same symmetry pattern, i.e. mainly localized along the positive $z$ axis. Figure \ref{Orbits}a shows a vibrational-type orbit. The two stable fixed points located symmetrically left and right from the central region corresponds to almost circular orbits with opposite angular momenta. The levels around these points are rotational-type quasiperiodic orbits with the same symmetry pattern. Figure \ref{Orbits}b shows a rotational-type orbit. Finally, the levels in the exterior region are vibrational-type quasiperiodic orbits mainly oriented along the negative $z$ axis.

Now we focus on the opposite regime, when the energy is high enough so that the electron can escape from the nucleus attraction. In this case, the system tends to be nonintegrable and the electron has access to the ionization channel located along the negative $z$ axis. In figures \ref{Poincare}c and \ref{Poincare}d we show Poincar\'{e} surfaces of section for $\xi = -1$ ($n = 71$) and $\xi = - 0.5$ ($n = 100$), respectively. A rich variety of behavior is observed in the Poincar\'{e} section for the system as the energy $\xi$ increase. In fig. \ref{Poincare}c we observe a similar behavior to that of the case $\xi < \xi _{c}$. In fig. \ref{Poincare}d we observe that the vibrational-type closed orbits oriented along the negative $z$ axis, representing quasiperiodic behavior, are replaced by irregular patterns, and eventually the Poincar\'{e} plane seems to be swamped by chaos. An interesting consequence of this system is the atomic ionization for $\xi > \xi _{c}$, which can be understood in the phase space as follows. Since the escape channel is located along the negative $z$ axis ($v$ axis in regularized coordinates), when $\xi > \xi _{c}$ the levels in the exterior region of the Poincar\'{e} surfaces of section are not bounded regions because the rectilinear orbits (along the $v$ axis with $u=0$) are the first orbits to ionize. Since not all trajectories ionize because part of the phase space remains isolated from the ionization channel (mainly the rectilinear orbits along the positive $z$ axis, or $u$ axis with $v=0$ in regularized coordinates), ionized orbits will appear in the phase space as unbounded regions enveloping stable orbits. This behavior is not shown in fig. \ref{Poincare}. A deeper understanding of this charge transfer mechanism is obtained when described in terms of chemical reaction dynamics. In this letter we focus on the consequences of axion electrodynamics in the quasiperiodic orbits, thus the problem of atomic ionization due to TI surfaces is left for future investigations.

\section{Tuning the dynamics of the atomic electron}

In the previous section we have considered the case of the TI TlBiSe$_{2}$, for which $\theta = \pi$, $\varepsilon = 4$ and $\mu = 1$. As discussed before, when TR symmetry is broken on the surface of the TI (by adding a magnetic coating) the system becomes fully gapped and the TMEP is quantized in odd integer values of $\pi$ such that $\theta = \pm (2n+1) \pi$, where $n \in \mathbb{Z}$. The two signs correspond to the two possible orientations of the magnetization in the direction perpendicular to the surface, and $2n+1$ corresponds to the number of surface fermions. As we will demonstrate in the following, the dynamical properties of the system are sensitive to the value of $\theta$, and the theoretical tunability of its value will allow us to describe a mechanism for switching between vibrational and rotational types of orbits.

\begin{figure}[tbp]
\begin{center}
\includegraphics[width=1.7in]{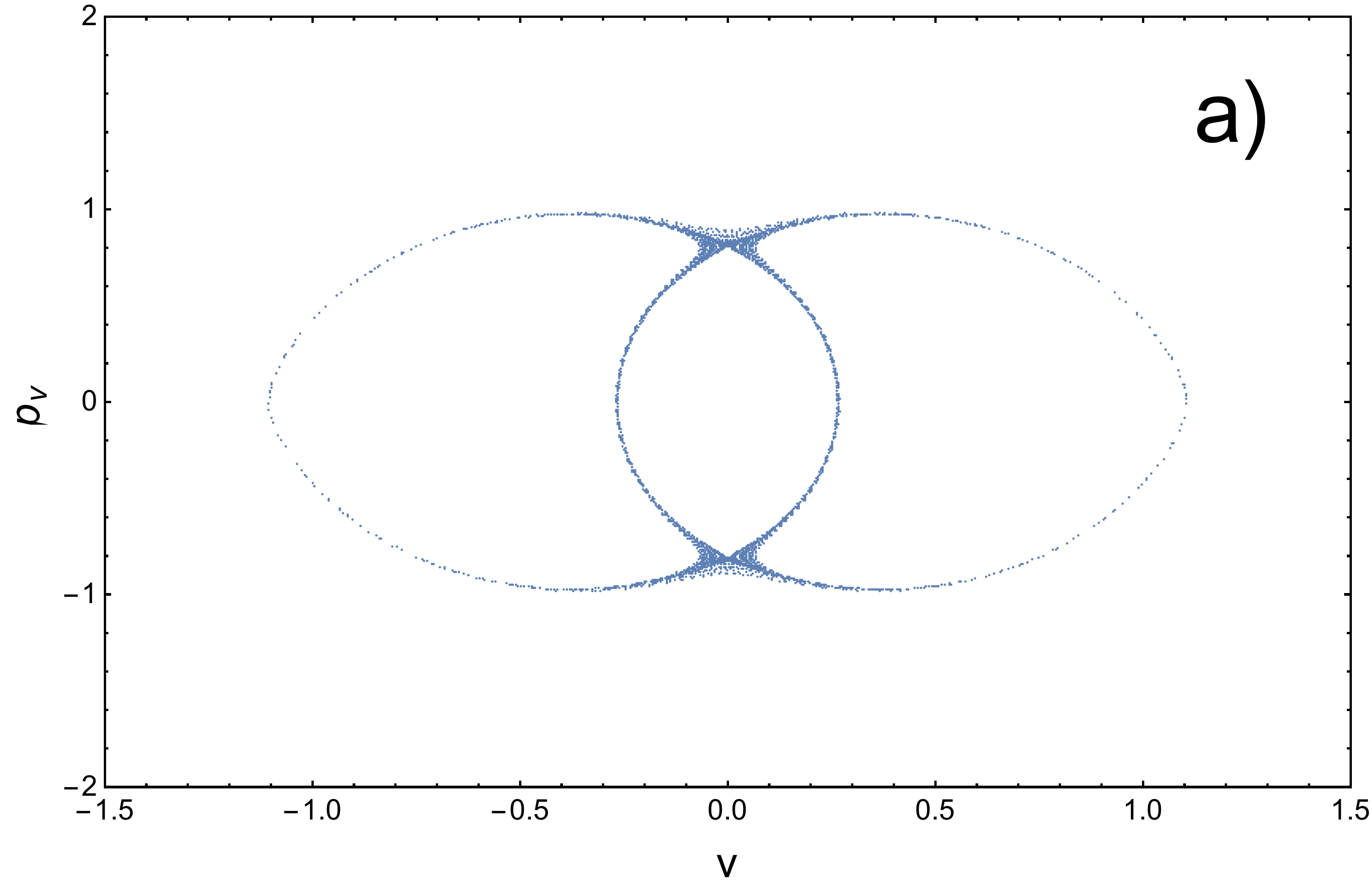} \label{PSE1}
\includegraphics[width=1.7in]{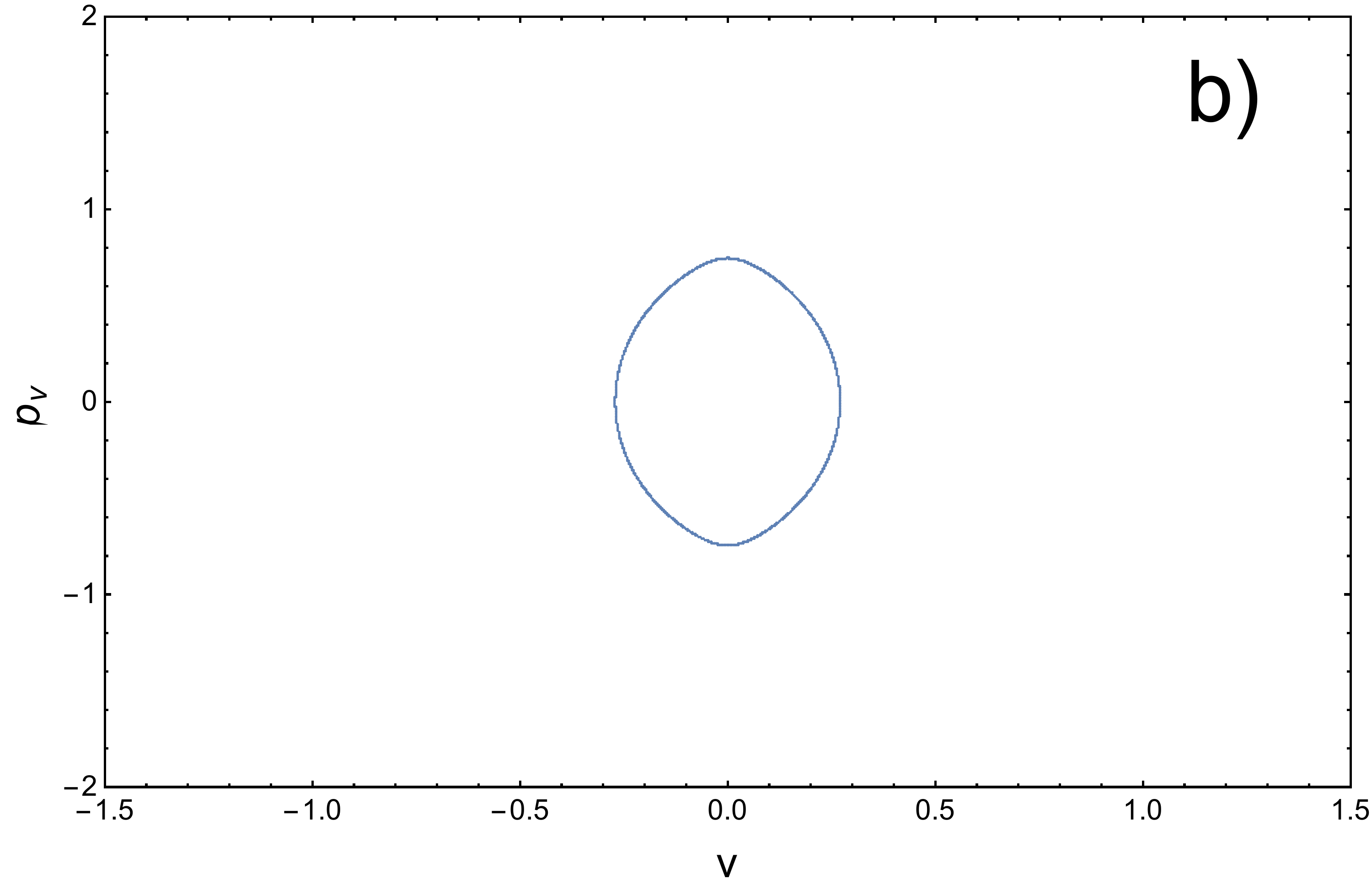} \label{PSE2} \\
\includegraphics[width=1.7in]{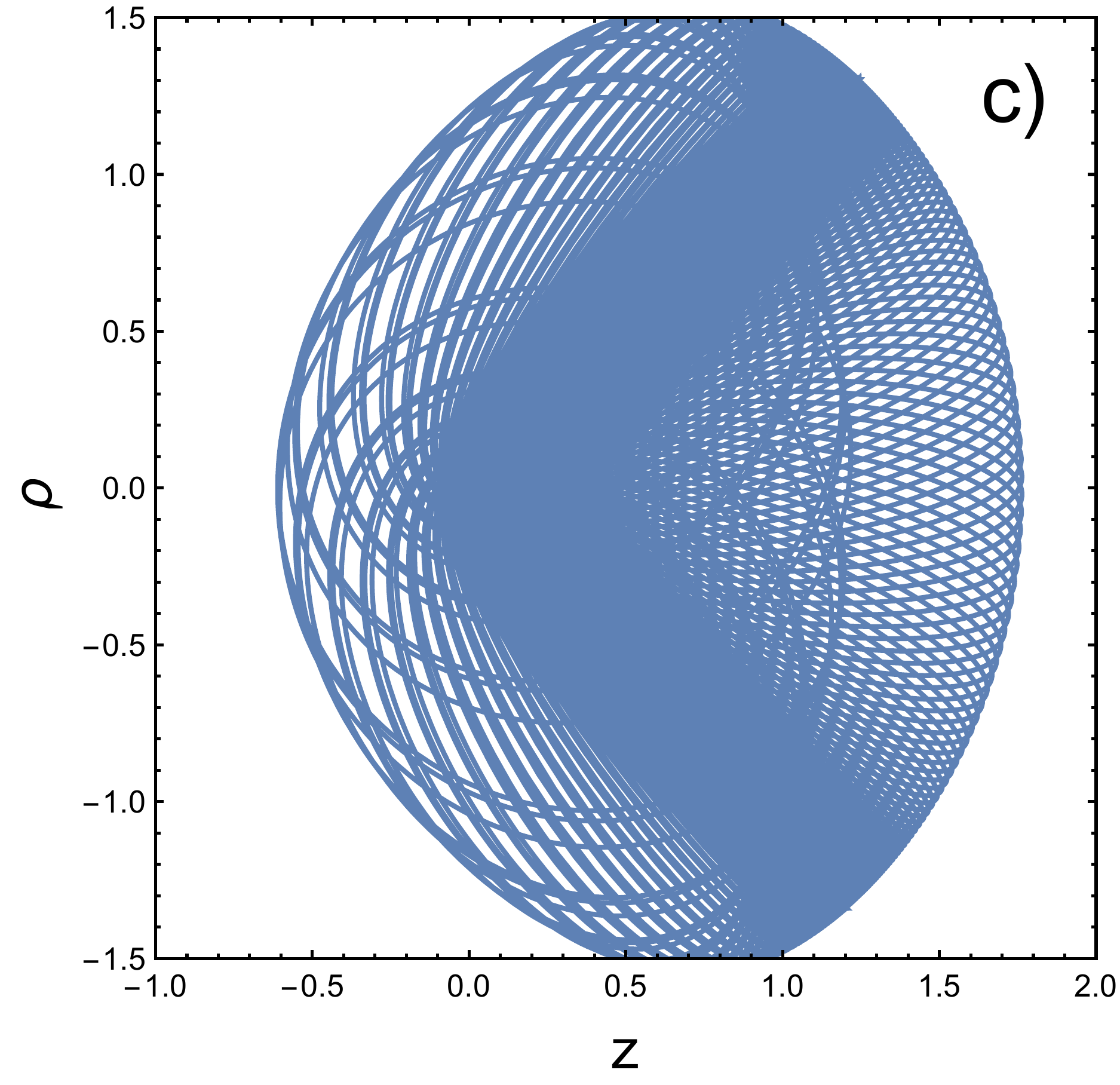} \label{RotE1}
\includegraphics[width=1.7in]{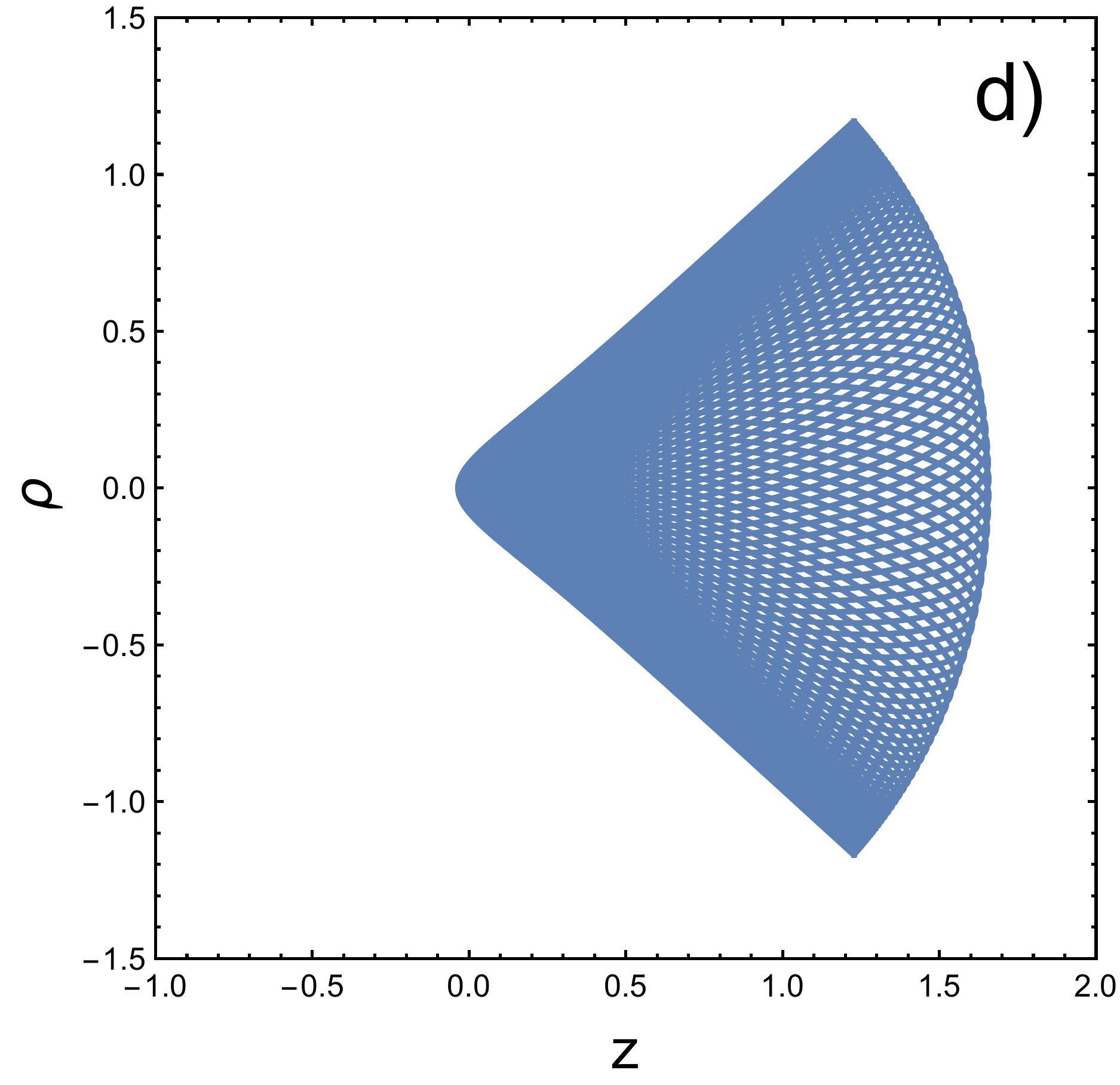} \label{VibE2}
\end{center}
\caption{{\protect\small Tuning the dynamics with the TMEP.}}
\label{Poincare2}
\end{figure}

From eqs. (\ref{kappa}) and (\ref{ImageMonopole}) we can see that high values of $\theta$ and low values of $\varepsilon$ favor the topological contribution. Therefore, to amplify the effects due to the topological nontriviality of the material, let us consider the value of $\theta$ which maximizes the magnetic monopole strength $g$, which is $\tilde{\alpha} _{c} = \pm \sqrt{(\varepsilon +1)(1 + 1 / \mu)}$. This critical value produces $\kappa = \varepsilon / (\varepsilon + 1)$ and $g = 1 / \tilde{\alpha} _{c}$. Now let us consider an hypothetical topological insulator with the same optical properties to that of the TI TlBiSe$_{2}$ ($\varepsilon =4$, $\mu = 1$) but hosting a large number of Dirac cones in its surface. Most of the TIs discovered up to date exhibit a single or a small number of surface fermions, and the problem of characterizing such materials with large $\theta$ is still an open question. Therefore, although in practice the tunability of the TMEP from zero to high values is very unlikely, it is suitable to illustrate the effects of the topological nontriviality. In order to analyze the consequences of switching the TMEP from zero to high values, let us first consider the case $\theta = 0$ and compute the Poincar\'{e} surface of section for a given initial condition $\boldsymbol{\lambda} _{0} (\xi) = (p _{u 0} , p _{v 0} , u _{0} , v _{0} )$ and scaled energy $\xi = - 0.5$, as shown in fig. \ref{Poincare2}a. Here we recall that one of the components of the initial condition (e.g. $p _{v 0}$) is determined from the other through the energy shell condition $\tilde{\mathcal{H}} ^{\prime} (\boldsymbol{\lambda} _{0}, \tilde{\alpha} = 0) = 2$. We conveniently choose the rotational-type of orbit depicted in fig. \ref{Poincare2}c. If we turn on the $\theta$-parameter from zero to $\pi$, the Kolmogorov-Arnold-Moser theorem guarantees that the orbits do not significantly change their character with respect to the case $\theta = 0$, since the perturbations to the Hamiltonian due to the topological contributions are very small (of the order of $\alpha$). However, if we turn on the $\tilde{\alpha}$ parameter to its critical value $\tilde{\alpha} _{c}$, such perturbations will not longer be small, but strong enough to produce significant changes in the electron dynamics (KAM theorem does not hold), as we will demonstrate in the following. Due to the energy shell condition $\tilde{\mathcal{H}} ^{\prime} (\boldsymbol{\lambda} _{1}, \tilde{\alpha} _{c}) = 2$, the initial condition now becomes $\boldsymbol{\lambda} _{1} (\xi) = (p _{u 0} , p _{v 1} , u _{0} , v _{0} )$. As shown in fig. \ref{Poincare2}b, the Poincar\'{e} surface of section significantly changes its character due to the strong perturbations. Clearly, a rotational-vibrational type transition is induced due to the change in $\tilde{\alpha}$. The corresponding vibrational-type orbit is shown in fig. \ref{Poincare2}d. One can also demonstrate that a vibrational-rotational type transition can also be tuned with the TMEP.

\section{Conclusions} \label{Conclusions}

The problem of the dynamics of Rydberg atoms near surfaces is important in its own in the fields of physics, chemistry and biology. When a slowly moving atom or ion approaches to a metallic surface, the mutual interaction leads to electronic processes of great interest in physics. For example, as the atom approaches the surface, the outer electron is captured by the surface and the atom ionizes. After this charge transfer process, the positive ion is attracted to the surface by its image charge and finally it is neutralized by an Auger process. This system has been widely studied in the literature and extensions to dielectric surfaces has also been analyzed. 

In this letter, we have studied the classical dynamics of a Rydberg hydrogen atom near the surface of a planar topological insulator, as shown in fig. \ref{System}. Due to the large size of a Rydberg atom, the interaction with the TI surface takes place relatively far from the surface and is dominated by nonretarded electromagnetic forces. By virtue of the TME, the electric charges of the atom produce image electric charges and image magnetic monopoles located inside the material, which in turn interact with the atomic electron thus affecting its dynamics. Owing to the axial symmetry of the system, when the Hamiltonian is expressed in cylindrical coordinates, the $z$ component $l _{z}$ of the angular momentum is conserved, and the system reduces to two degrees of freedom. We have restricted our analysis to the case $l _{z} = 0$.

By means of numerical techniques and Poincar\'{e} surfaces of section, we have explored the phase space structure of the system for the TI TlBiSe$_{2}$. The corresponding surfaces of section are shown in fig. \ref{Poincare} for different values of the scaled energy $\xi$, where we have distinguished important structures which include stabled fixed points. The levels around these points include vibrational (fig. \ref{Orbits}a) and rotational (fig. \ref{Orbits}b) types of orbits. When the energy of the electron is bigger than the critical energy, the electron has access to the ionization channel and can be captured by the TI surface.

Interestingly, we have shown that vibrational-rotational-vibrational type transitions can be tuned with the TMEP $\theta$. To this end, we first consider a particular rotational-type orbit of the atomic electron for $\theta = 0$. Next we turn on the $\theta$-parameter to its critical value $\tilde{\alpha} _{c}$ and we observe the transition to a vibrational-type orbit. We provide a visual demonstration of this effect in fig. \ref{Poincare2}. Importantly, this transition is a unique signature of the topological nontriviality of the material. In practice, to observe this effect we require a topological insulator hosting a large number of surface fermions in its boundary; however the problem of characterizing such materials is still an open question in condensed matter physics. The proposed effect could also be explored in other magnetodielectric materials which are described by higher axion couplings, such as Cr$_{2}$O$_{3}$ \cite{Cr2O3}. However, these materials induce more general magnetoelectric couplings not considered in our model \cite{Essin}.

A further extension of this work is the full quantum-mechanical treatment of the system \cite{Urrutia}. Also, inspired by the recent discovered semimetallic phases of the quantum matter, it would be interesting the analysis of the dynamics of Rydberg atoms near Dirac and Weyl semimetals. We hope that our results will be useful in understanding the dynamical behavior of atoms, ions, or molecules near topological phases of matter, and encourage experimental efforts to attain full characterization of these novel phases of matter.

\acknowledgments

AMR has been supported in part by the projects DGAPA(UNAM) IN104815 and CONACyT (M\'{e}xico) 237503.

\end{document}